*Title*

# Nanofocusing optics for an X-ray free-electron laser generating an extreme intensity of 100 EW/cm² using total reflection mirrors

**Hirokatsu Yumoto [1,2,\*], Yuichi Inubushi [1,2], Taito Osaka [2], Ichiro Inoue [2], Takahisa Koyama [1,2], Kensuke Tono [1,2], Makina Yabashi [1,2] and Haruhiko Ohashi [1,2]**

1. Japan Synchrotron Radiation Research Institute, 1-1-1, Kouto, Sayo-cho, Sayo-gun, Hyogo 679-5198 Japan; inubushi@spring8.or.jp (Y.I.); koyama@spring8.or.jp (T.K.); tono@spring8.or.jp (K.T.); hohashi@spring8.or.jp (H.O.)
2. RIKEN SPring-8 Center, 1-1-1 Kouto, Sayo-cho, Sayo-gun, Hyogo 679-5148 Japan; osaka@spring8.or.jp (T.O.); inoue@spring8.or.jp (I.I.); yabashi@spring8.or.jp (M.Y.)
\* Correspondence: yumoto@spring8.or.jp; Tel.: +81-791-58-0802

**Abstract:** A nanofocusing optical system referred to as *100 exa* for an X-ray free-electron laser (XFEL) was developed to generate an extremely high intensity of 100 EW/cm² ($10^{20}$ W/cm²) using total reflection mirrors. The system is based on Kirkpatrick-Baez geometry, with 250 mm long elliptically figured mirrors optimized for the SPring-8 Angstrom Compact Free-Electron Laser (SACLA) XFEL facility. The nano-precision surface employed is coated with rhodium and offers a high reflectivity of 80%, with a photon energy of up to 12 keV, under total reflection conditions. Incident X-rays on the optics are reflected with a large spatial acceptance of over 900 μm. The focused beam is 210 nm × 120 nm (full width at half maximum) and was evaluated at a photon energy of 10 keV. The optics developed for *100 exa* efficiently achieved an intensity of $1 \times 10^{20}$ W/cm² with a pulse duration of 7 fs and a pulse energy of 150 μJ (25% of the pulse energy generated at the light source). The experimental chamber, which can provide varied stage arrangements and sample conditions, including vacuum environments and atmospheric pressure helium, was set up with the focusing optics to meet the experimental requirements.

## 1. Introduction

Hard X-ray free-electron laser (XFEL) sources [1–5] have successfully generated extremely high-peak brilliance unachieved by existing synchrotron radiation sources. Focusing optics to further enhance the pulse intensity is a particularly useful tool for XFEL experiments. There have been several reports on hard XFEL focusing optics for producing high-intensity pulses utilizing diffractive [6,7], reflective [8,9], and refractive [10,11] optics. Among these alternatives, total reflection optics possess remarkable properties in terms of achromaticity, high efficiency, and high tolerance under intense XFEL irradiation.

In a previous study, we developed a two-stage focusing optics system [9] that achieves a 50 nm focus size and an extremely high intensity of $1 \times 10^{20}$ W/cm² at the SPring-8 Angstrom Compact Free-Electron Laser (SACLA) XFEL facility [2]. This optics system utilized nano-precision figured mirrors with an elliptical cylinder shape in Kirkpatrick–Baez (K–B) geometry [12]. Although two-stage focusing optics have successfully contributed to the observation of nonlinear X-ray optical phenomena [13–16], we encountered several practical challenges in the operation of this system (such as stability, repeatability, and alignment) because the focusing condition of the system is highly sensitive to mirror misalignment. Furthermore, the optics system had a low throughput of less than 10% because it had a small spatial acceptance.

In this study, we report the development of a highly stable, high-throughput focusing system that achieves a high intensity of $1 \times 10^{20}$ W/cm² (comparable to that of our previous two-stage focusing system) by utilizing single-stage K–B optics optimized for BL3 of SACLA. The system's large spatial acceptance mirrors of over 900 μm enabled a threefold increase in pulse energy of 150 μJ relative to the previous system, a high intensity of $1 \times 10^{20}$ W/cm² at a photon energy of 10 keV, and a highly stable beam for over 13.5 hours while achieving a focal spot size of approximately 210 nm × 120 nm (full width at half maximum (FWHM)).

## 2. Design

### 2.1. Optics

The optical parameters of the elliptical cylinder shaped focusing mirrors designed for the optical system are summarized in Table 1. The focusing system has a wide spatial acceptance of 970 μm and 920 μm in the vertical and horizontal directions, respectively, which is considerably larger than the incident beam size of 450 μm (FWHM) at EH5 of BL3 at a photon energy of 10 keV. The designed K–B optics have an aperture receiving 80% of the incident pulse energy at a photon energy of 10 keV. We designed the working distance, (length between the downstream edge of the optics and the focus), to be 115 mm, which is long enough to conduct various experiments. The focused beam sizes were geometrically calculated to be 182 and 87 nm (FWHM) using a demagnification ratio based on the light source size and the focal lengths. The focus size is broader than that of the two-stage focusing system because of the moderate demagnification factor. The calculated focus sizes are three times larger than the diffraction-limited beam sizes of 56 and 28 nm (FWHM) determined using the numerical aperture of the optical system. These optics are advantageous for maintaining the optimized focusing condition because the alignment tolerances of both the incident angle and the focal length are approximately 10 times larger than those of the previous two-stage focusing optics system.

**Table 1.** Optical parameters of the focusing mirrors

|  | Vertical focusing mirror | Horizontal focusing mirror |
|---|---|---|
| Surface profile | Elliptical cylinder | Elliptical cylinder |
| Substrate material | Single crystal silicon | Single crystal silicon |
| Surface coating | Rhodium | Rhodium |
| Mirror substrate size | 250 × 50 × 50 mm³ | 250 × 50 × 50 mm³ |
| Glancing angle on optical axis | 4.0 mrad | 3.8 mrad |
| Focal length | 500 mm | 240 mm |
| Semimajor axis | 110.25 m | 110.25 m |
| Semiminor axis | 41.95224 mm | 27.62842 mm |
| Diffraction-limited focus size in FWHM at 10 keV | 56 nm | 28 nm |
| Effective mirror length | 242 mm | 242 mm |
| Spatial acceptance | 970 μm | 920 μm |

The incident angles of the mirrors were designed for use under total reflection conditions of up to 12 keV. Figure 1(a) presents the calculated reflectivity of the mirrors plotted against the incident photon energy. The reflectivity curves for a rhodium surface coating with a density that is 100% of the bulk density are presented in Figure 1(b) and (c). The reflectivity of the two mirrors is calculated to be 80% up to a photon energy of 12 keV, and above 50% at a photon energy of 15 keV. The rhodium-coated mirrors are free from absorption edge within a photon energy range of 5–15 keV. The radiation dose to the rhodium in the optical system is estimated to be well below the threshold level for radiation damage from intense XFEL [17]. With these design parameters, the pulse energy, fluence, and intensity of the focus were calculated to be ~200 µJ, $2 \times 10^6$ J/cm², and $2 \times 10^{20}$ W/cm² at a photon energy of 10 keV, when the incident pulse energy to the optics and the pulse duration were assumed to be 300 µJ and 7 fs.

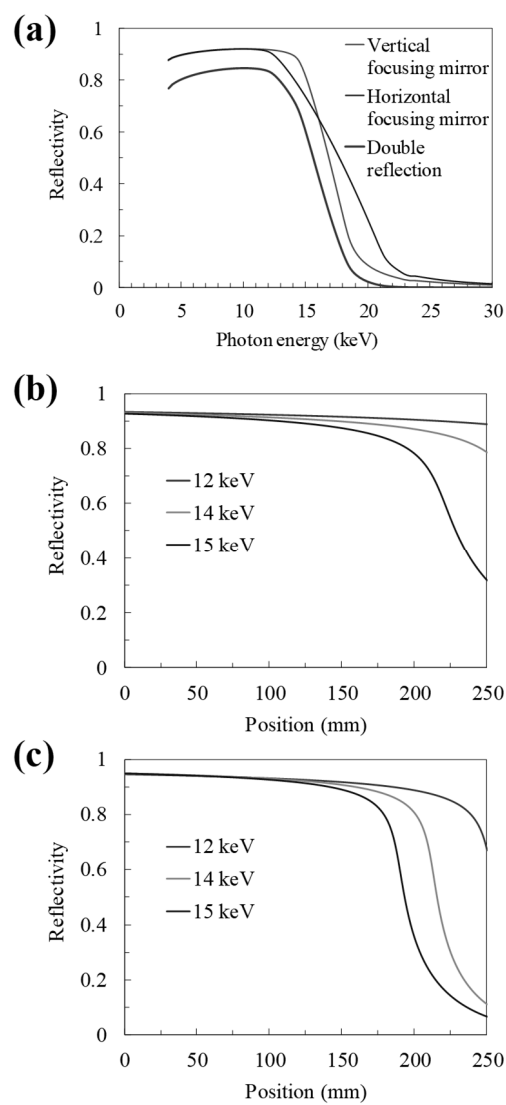

**Figure 1.** Reflectivity of designed mirrors. (a) Dependence of reflectivity on the incident photon energy. (b) Dependence of reflectivity on the surface position of the vertical focusing mirrors and (c) the horizontal focusing mirrors. Position 0 mm shows the upstream edge of the focusing mirror. The change in reflectivity is due to variation in the incident angle along the elliptical cylinder surface.

*2.2. Mirrors and sample chambers*

The apparatus developed consists of two independent vacuum chambers: one for the mirrors and the other for the sample. The vacuum chambers are separated by a beryllium window, which protects the mirror surfaces from contaminants from the sample chamber, including particles produced via the ablation of sample materials. The mirror chamber was designed such that the mirrors can be operated under high vacuum conditions below $10^{-5}$ Pa. In the mirror chamber, we installed stable, high-precision alignment mechanics. Greaseless driving mechanisms with stepper motors are used in all alignment stages of the mirrors, including a linear translational stage utilizing a solid lubricant. The rotating stages, which are composed of a precision linear actuator, with a precision of 10 nm/step (full-step drive) and a flexure hinge at the rotation center, are employed in the alignment of incident angles and perpendicularity between the mirrors. These stages are operated at a precision higher than 0.1 μrad using a harmonic drive gear and a micro-step drive control, while the driving mechanics of the incident angles need to be stabilized at a level of 0.1 μrad to obtain high pointing stability of the optical system.

The sample chamber has an effective space of over 100 mm in length from the beryllium window to the focal point, as shown in Figure 2. The typical vacuum pressure of this chamber is 0.1–1 Pa, and it can be evacuated using a turbo-molecular pump. The sample environment can also be operated under atmospheric pressure helium conditions depending on the sample requirements of the experiments. Various sample stages and sample supply systems, including a liquid jet injector, can be placed in the space around the focal point.

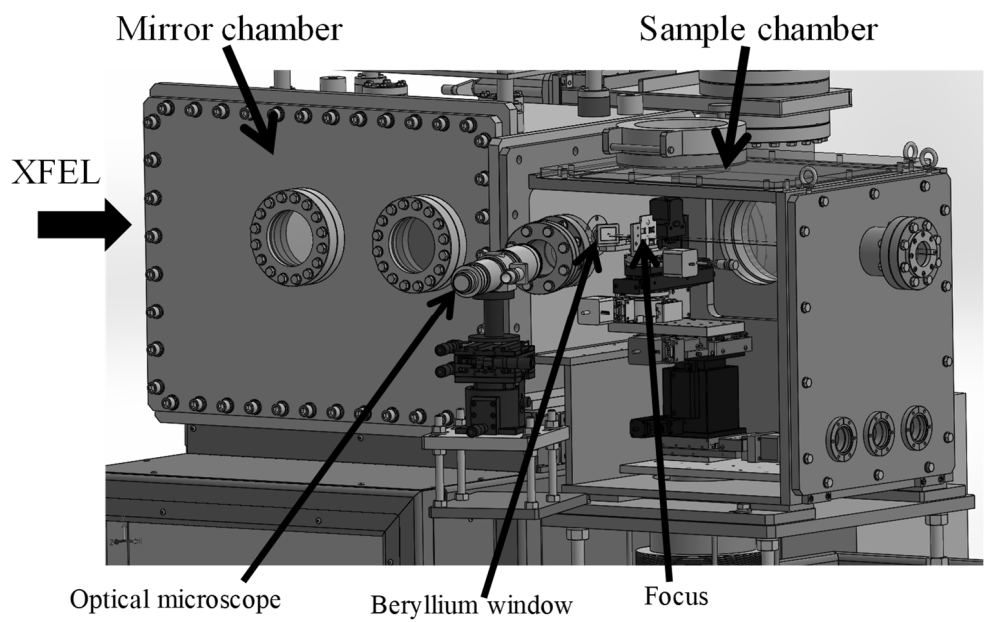

**Figure 2.** Schematic drawing of the mirror and the sample chamber. The wall on the near side of the sample chamber is presented as transparent to show the inside of the chamber. A beryllium window separates the mirror and sample chambers. An optical microscope is used to observe the sample at the focus from outside the vacuum chamber. Various sample environments, stage arrangements, and sample supply systems can be operated utilizing a 100 mm space upstream of the focus. A portion of the sample chamber can be altered based on experimental requests while keeping the mirror chamber unchanged.

## 3. Evaluations at SACLA

### 3.1. Evaluation of focusing optics

The designed mirrors were fabricated using an ultra-precise computer-controlled process [18] with a surface figure precision (figure error) of 0.30 to 0.52 nm root mean square (RMS) and surface roughness of 0.11 to 0.14 nm RMS. The properties of the focused beam were evaluated at EH5, BL3 of SACLA. The focusing mirrors were aligned using a Foucault knife-edge test [19], with a 200 μm diameter gold wire. For focusing a beam size of 100 nm (FWHM), the alignment tolerances of the horizontal mirror (which are more severe than those of the vertical mirror) were estimated to be ±1.6 μrad in the incident angles, ±10 μrad in perpendicularity between the mirrors, and ±2 mm in the focal length at a relatively high photon energy of 12 keV.

Intensity distributions of the focus were measured via the knife-edge scanning method using the wire. The typical focused beam profiles obtained at a photon energy of 10 keV are presented in Figure 3. To prevent ablation of the wire by the intense focused beam, the incident pulse energy for the mirrors was adequately attenuated with silicon attenuation plates. The shot-to-shot fluctuations of the incident pulse energy to the mirrors were measured with an intensity monitor [20] and were normalized to eliminate their influence on the profile measurements. The typical beam sizes observed, 210 nm × 120 nm (FWHM), in the vertical and horizontal directions, show most of the reflected pulse energy concentrated at the main peak and low-intensity distributions around the main peak. The focused beam had a pulse energy of 150 μJ, which is 25% of the pulse energy generated at the XFEL light source. From these measured values, the fluence and intensity of the focused pulses reached $8 \times 10^5$ J/cm$^2$ and $1 \times 10^{20}$ W/cm$^2$, assuming a pulse duration of 7 fs [21,22].

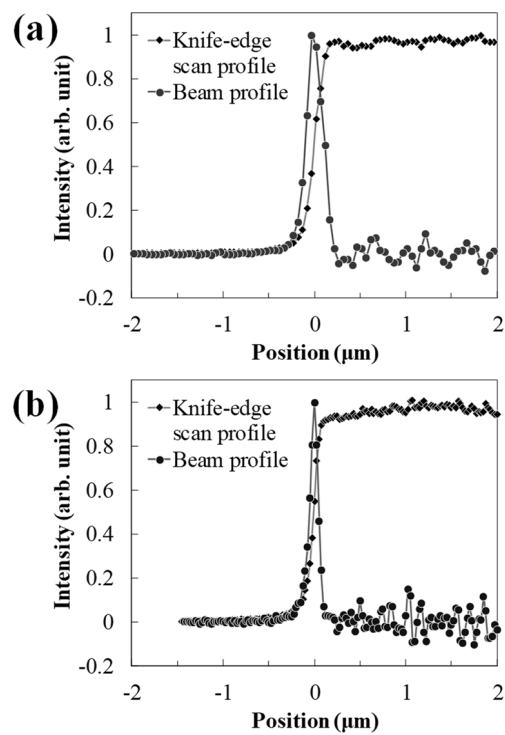

**Figure 3.** Typical intensity distributions of focused beam at a photon energy of 10 keV measured using the knife-edge scanning method. The focused beam size of 210 nm × 120 nm (FWHM) was evaluated in the (a) vertical direction and (b) horizontal direction.

*3.2. Stability of the focused beam*

The stability of the focused beam was assessed in terms of size, position, and profile, as shown in Figure 4. The shot-to-shot positional displacement of the focused pulses was less than the measured beam size because the profile measured using the knife-edge scanning method is an average of many pulses. As shown in Figure 4(a), (b), and (c), the beam sizes and profiles in both the vertical and horizontal directions were maintained for over 13.5 hours after the alignment of the mirrors. Regarding the focus position in the same evaluation, the positional displacement observed was about 2.5 µm and 0.4 µm in the vertical and horizontal directions, as shown in Figure 4(b) and (c). The evaluation results for the focal size, position, and profile show sufficient stability for most of the XFEL experiments in which a sample is destroyed via ablation using an extremely intense single pulse.

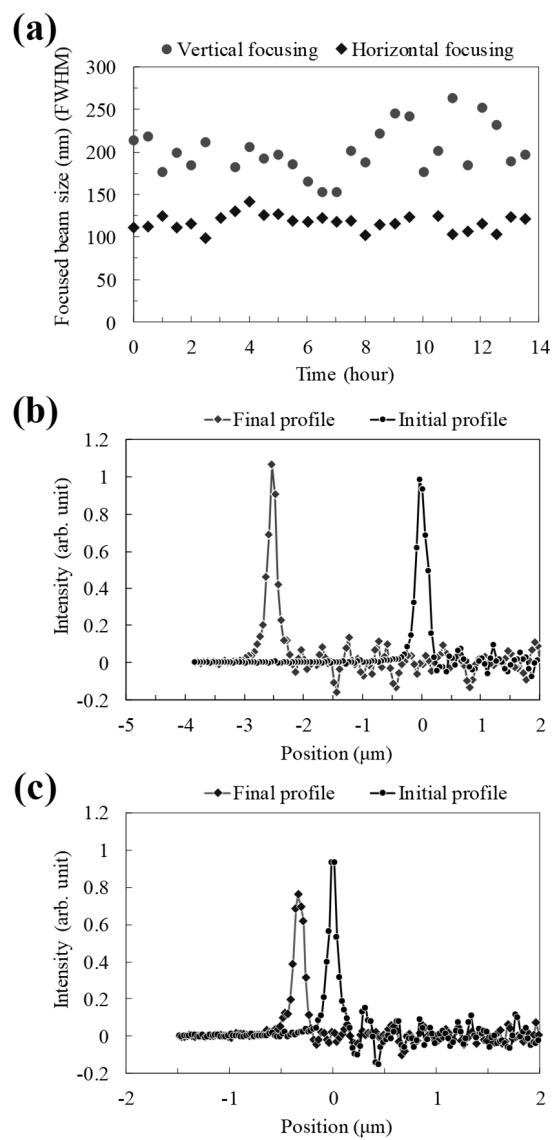

**Figure 4.** Measured stability of focused beam. (a) Time dependence of focused beam size. (b) Focused beam profiles observed during stability evaluation, in the vertical direction, and (c) horizontal direction.

## 4. Conclusion

We demonstrate a highly-efficient single-stage K–B optics that generates intense XFEL focused pulses. The focused beam evaluated achieved a high fluence of $8 \times 10^5$ J/cm$^2$ and an intensity of $1 \times 10^{20}$ W/cm$^2$, with a significantly improved pulse energy of 150 μJ and stable beam size. Recently, nanofocusing optics for extremely high intensities of $10^{21}$ W/cm$^2$ have been reported [7], with multilayer mirrors having a narrow bandpass. In contrast, the optics system in this study, with total reflection mirrors, is applicable for a variety of XFEL experiments, such as two-color double-pulse experiments [23,24], which require extremely high intensity, achromaticity, and highly improved stability.

**Author Contributions:** Optical design of the focusing mirrors: H.Y., Y.I., T.K., and H.O. Design of the mirror alignment mechanism: H.Y., T.K., and H.O. Sample chamber design: Y.I., T.O., I.I., K.T., and M.Y. Evaluation of the focusing beam: Y.I., T.O., and I.I. Writing—original draft preparation: H.Y. Writing—review and editing: M.Y. Project administration: M.Y. All authors have read and agreed to the final version of the manuscript.

**Acknowledgments:** We would like to acknowledge the supporting members of SACLA facility. The XFEL experiments were performed at the BL3 of SACLA with the approval of the Japan Synchrotron Radiation Research Institute (JASRI) (Proposal No. 2017B8086).